# Unified Growth Theory Contradicted by the Economic Growth in Latin America

Ron W Nielsen[1]

Environmental Futures Research Institute, Gold Coast Campus, Griffith University, Qld, 4222, Australia

January, 2016

Historical economic growth in Latin America is analysed using the data of Maddison. Unified Growth Theory is found to be contradicted by these data in the same way as it is contradicted by the economic growth in Africa, Asia, former USSR, Western Europe, Eastern Europe and by the world economic growth. Paradoxically, Unified Growth Theory is repeatedly and consistently contradicted by the same data, which were used, but never properly analysed, during the formulation of this theory. Unified Growth Theory does not explain the mechanism of the economic growth because it explains features contradicted by data. This theory is based fundamentally on the unfortunate lack of understanding of the properties of hyperbolic distribution and on the unscientific analysis of data. There was no transition from stagnation to growth at the end of the alleged Malthusian regime because the economic growth was hyperbolic. There was no escape from Malthusian trap because there was no trap. There was no takeoff. On the contrary, at the time of the alleged takeoff economic growth started to be diverted to a slower trajectory. Unified Growth Theory is dissociated from the reality. This theory needs to be revised or replaced. In its present form, it is a collection of irrelevant stories based on impressions and on the unscientific use of data.

**Introduction**

We have already demonstrated that the Unified Growth Theory (Galor, 2005a; 2011) is contradicted by the economic growth in Africa, Asia, former USSR, Western Europe and Eastern Europe as well as by the world economic growth (Nielsen, 2014, 2015a, 2015b, 2015c, 2015d, 2015e). Now, we shall show that this theory is also contradicted by the economic growth in Latin America.

In our analysis we shall use the latest data published by Maddison (2010). Galor used the earlier publication (Maddison, 2001) but any of these publications can be used to show that the Unified Growth Theory is dramatically contradicted by Maddison's data.

Again we shall show that, paradoxically, Unified Growth Theory is *contradicted by the same data, which were used during its development*. Such a paradox is hard to find in science, but it is common in discussions of doctrines accepted by faith. Galor appears to have been guided by the accepted doctrines but he also appears to have been experiencing a genuine problem with the analysis of data. His irrelevant explanations of the mechanism of economic growth are based on a meaningless quotations of isolated numbers, on the unfortunate inexperienced inspection of data and on the habitual use of grossly distorted diagrams (Ashraf, 2009; Galor, 2005a, 2005b, 2007, 2008a, 2008b, 2008c, 2010, 2011, 2012a, 2012b, 2012c; Galor and Moav, 2002; Snowdon & Galor, 2008).

---

[1] AKA Jan Nurzynski, r.nielsen@griffith.edu.au; ronwnielsen@gmail.com; http://home.iprimus.com.au/nielsens/ronnielsen.html

Data are essential in scientific research and they have to be treated with care and respect. Such a treatment of data as repeatedly manifested in the Unified Growth Theory and other associated publications can never lead to reliable conclusions. The ironic feature of this theory is that the data appear to be correct and reliable (not all of them have been checked) but their interpretation is certainly incorrect because it is *not* based on their scientific and rigorous analysis.

Historical economic growth, represented by the Gross Domestic Product (GDP), is hyperbolic (Nielsen, 2015f). These types of distributions can be misleading even for the most experienced researcher because they create an illusion of being made of two distinctly different components, slow and fast, while in fact they represent a single, monotonically-increasing trajectory. Fortunately, the analysis of these distributions is trivially simple (Nielsen, 2014). The GDP per capita (GDP/cap) distributions are even more deceptive but their analysis is also relatively simple (Nielsen, 2015a). When properly analysed, these distributions also demonstrate that the Unified Growth Theory is contradicted by the data describing economic growth (Nielsen, 2015a).

Latin America is made of less-developed countries (BBC, 2014; Pereira, 2011). According to Galor (2005a, 2008a, 2011, 2012a), economic growth in these countries should have been characterised by two distinctly different regimes of growth: the Malthusian regime of stagnation, which supposed to have ended around 1900 and the post-Malthusian regime, which allegedly commenced around that year. For these countries, the Industrial Revolution, 1760-1840 (Floud & McCloskey, 1994), was still within the Malthusian regime of stagnation. Galor's two regimes were supposed to have been governed by distinctly different mechanisms of growth and the transition from stagnation to growth was supposed to have been marked by a *takeoff*, the signature so clear and so distinct that it cannot be missed. Indeed, Galor describes it as a "remarkable" or "stunning" escape from Malthusian trap (Galor, 2005a, pp. 177, 220).

We shall demonstrate that Galor's story is contradicted remarkably by data. We shall show that the "remarkable" and "stunning" escape from Malthusian trap did not happen because there was no need for any escape. There was no takeoff from stagnation to growth because the growth was not stagnant but hyperbolic. We shall demonstrate that at the time of this postulated remarkable takeoff, economic growth started to be diverted to a slower trajectory.

We shall show that Galor's account of the economic growth and its mechanism is contradicted by the data for Latin America in much the same way as it is contradicted by the economic growth in Africa, Asia, former USSR, Western Europe and Eastern Europe (Nielsen, 2014, 2015a, 2015b, 2015c, 2015d, 2015e) and by the world economic growth (Nielsen, 2014, 2015a). We shall demonstrate that the pattern of economic growth in Latin America is clearly different than the pattern claimed by the Unified Growth Theory.

As in our earlier publications, we shall use two ways of displaying data: (1) semilogarithmic display of the GDP data and (2) the display of their reciprocal values.

Again we shall remind that the hyperbolic growth is described by the simple mathematical formula:

$$S(t) = (a - kt)^{-1} \qquad (1)$$

where, in our case, $S(t)$ is the GDP while $a$ and $k$ are positive constants.

The reciprocal of a hyperbolic distribution is a straight line:

$$\frac{1}{S(t)} = a - kt \qquad (2)$$

If the reciprocal values of data follow a decreasing straight line, the growth is not stagnant but hyperbolic. However, the concept of stagnation is not supported even if the reciprocal values of data do not decrease linearly. Any monotonically-decreasing trajectory will show that the postulate of stagnation followed by a takeoff at the certain time is not supported by data. To prove the existence of the epoch of stagnation it is necessary to prove the presence of random fluctuations often described as Malthusian oscillations. Such random fluctuations should be clearly seen not only in the direct display of data but also in the display of their reciprocal values. It they are absent then there is no support in data for claiming the existence of the epoch of stagnation. Furthermore, if data do not show a clear takeoff from stagnation to growth at the postulated time, then there is no support for Galor's repeatedly-claimed takeoffs. However, if the reciprocal values of data follow a decreasing straight line, then they show, or at least strongly suggest, that the growth was hyperbolic.

If the straight line representing the reciprocal values of data remains unchanged, then obviously there is no change in the mechanism of growth. It makes no sense to divide a straight line into two or three arbitrarily selected sections and claim different regimes of growth controlled by different mechanisms for these arbitrarily-selected sections.

According to Galor, the transition from stagnation to growth occurred at the end of the alleged Malthusian regime and was marked by a clear and strong takeoff. For Latin America, the region made of less-developed countries (BBC, 2014; Pereira, 2011), this remarkable takeoff should have occurred around 1900 (Galor, 2005a, 2008a, 2011, 2012a). Such a takeoff should be clearly detected in the direct display of the GDP data but it should be seen even more clearly in the display of their reciprocal values and it should be indicated by a prominent *downward* change in the trajectory of the reciprocal values. We shall show that this prominent signature is missing in the data describing economic growth in Latin America. We shall show that the data tell one story while the Unified Growth Theory tells another and diametrically opposite story.

**Analysis of data for Latin America**

Results of analysis of the economic growth in Latin America based on Maddison's data (Maddison, 2010) are shown in Figures 1 and 2.

The data suggest the existence of two hyperbolic growth trajectories: a slow trajectory between AD 1 and 1500 and a fast trajectory between AD 1600 and 1870. The slow trajectory is characterised by parameters $a = 4.421 \times 10^{-1}$ and $k = 2.093 \times 10^{-4}$. The singularity for this trajectory was at $t_s = 2113$. The fast trajectory is characterised by parameters $a = 1.570 \times 10^0$ and $k = 8.224 \times 10^{-4}$. The singularity for this new trajectory was at $t_s = 1910$. However, from around 1870, i.e. from around the time of the alleged takeoff, the economic growth in Latin America started to be diverted to a slower trajectory bypassing the singularity by a safe margin of 40 years. The illusion of a takeoff is replaced by a diversion to a slower growth.

The characteristic features of the economic growth in Latin America are similar to the features observed for the economic growth in Africa, Australia and in Western Offshoots (Nielsen, 2015f). In all these examples, a slow hyperbolic growth was followed by a much faster hyperbolic trajectory and the transitions form slow to fast growths can be correlated with the intensified colonisation of the respective regions with the benefits of the intensified economic growth going to the colonising forces.

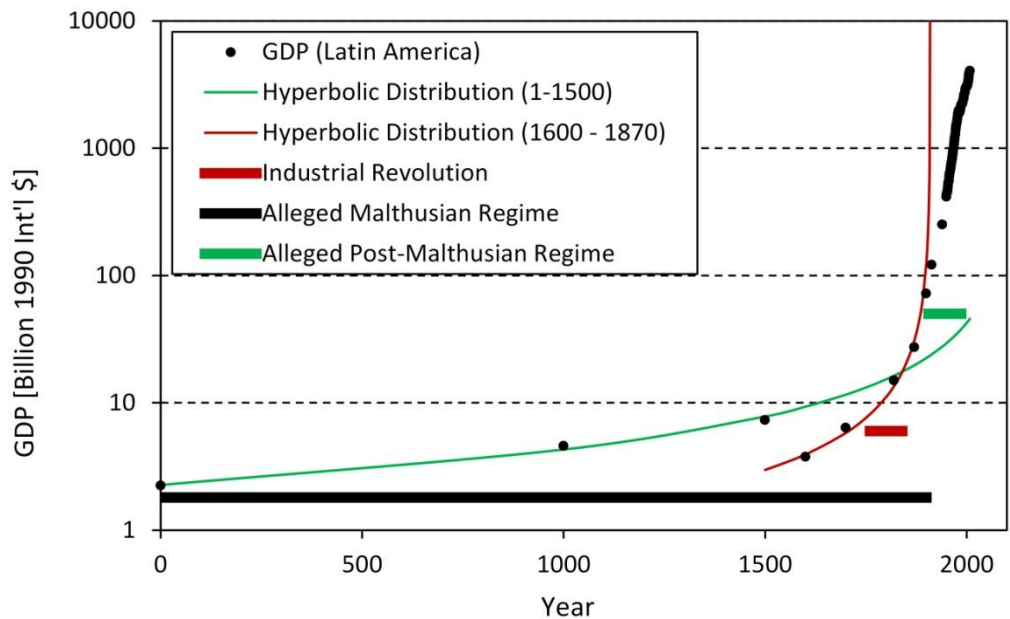

**Figure 1.** Economic growth in Latin America between AD 1 and 2008. The Gross Domestic Product (GDP) is in billions of 1990 International Geary-Khamis dollars. Maddison's data (Maddison, 2010) are compared with hyperbolic distributions and with their unsubstantiated interpretations proposed by Galor (2005a, 2008a, 2011, 2012a). The data suggest two hyperbolic distributions, the pattern similar to the economic growth in Africa (Nielsen, 2015b). The alleged transition from stagnation to growth never happened because the economic growth was not stagnant but hyperbolic. *Around the time of the postulated by Galor spectacular takeoff (around AD 1900) the economic growth started to be diverted to a slower trajectory*. There was no escape from Malthusian trap because there was no trap.

The data for Latin America are in clear disagreement with the Unified Growth Theory. The economic growth was slow before AD 1500 but there is no basis for claiming that it was stagnant. Hyperbolic trajectory between AD 1 and 1500 could be questioned but it is consistent with the similar, but much clearer, pattern in Africa and is in perfect agreement with the repeated evidence of hyperbolic growth in other regions (Nielsen, 2015f). However, in any case, there is definitely no convincing support in the data for the existence of the epoch of stagnation. We might imagine that there was stagnation but science is not built on imaginations alone. We might be wishing for a convincing evidence of the existence of the epoch of stagnation because this would be in harmony with the accepted doctrines and beliefs but the reality has no obligation to comply with our wishes and dogmas.

The data show a brief economic decline between AD 1500 and 1600, which appears to be coinciding with the commencement of the intensified Spanish conquest (Bethell, 1984). A similar brief decline is also demonstrated by the data for Australia (Nielsen, 2015f). However, from around AD 1600, economic growth in Latin America was following a fast-increasing hyperbolic trajectory. The change from a slow to fast economic growth occurred far too early for the Unified Growth Theory, about *300 years before the expected takeoff, which never happened*. Furthermore, as in Africa, *it was not a transition from stagnation to growth but from hyperbolic growth to hyperbolic growth*. This feature is ignored in the Unified Growth Theory. The theory presents a story, which is contradicted by data. There is no correlation between the data and the narrative of the Unified Growth Theory.

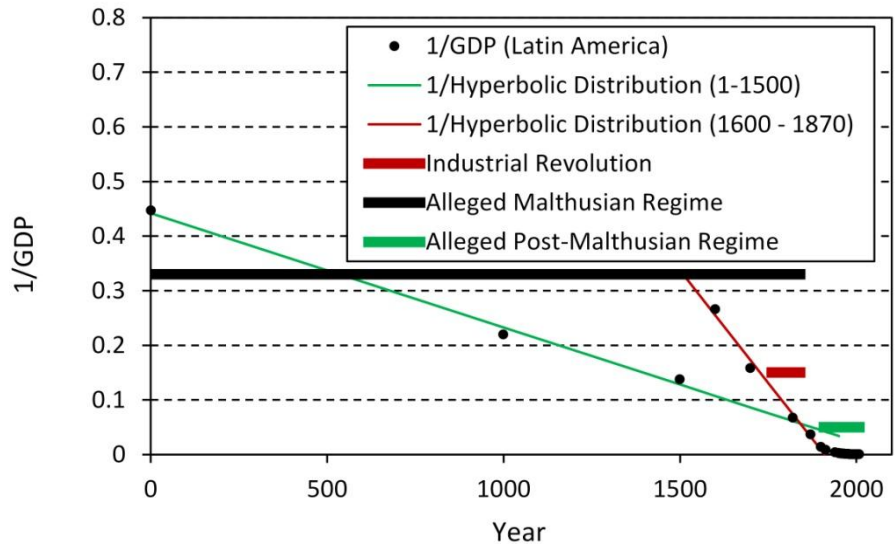

**Figure 2.** The reciprocal values of the Gross Domestic Product, 1/GDP, for Latin America between AD 1 and 2008. The GDP is in billions of 1990 International Geary-Khamis dollars. Maddison's data (Maddison, 2010) are compared with hyperbolic distributions represented by the decreasing straight lines and with their unsubstantiated interpretations proposed by Galor (2005a, 2008a, 2011, 2012a). The data suggest two hyperbolic distributions, the pattern similar to the economic growth in Africa (Nielsen, 2015b). The alleged transition from stagnation to growth never happened because the economic growth was not stagnant but hyperbolic. *Around the time of the alleged spectacular takeoff, the growth started to be diverted to a slower trajectory* as indicated by the *upward* bending of the trajectory of the reciprocal values. There was no escape from Malthusian trap because there was no trap. The transition from a slow to fast growth occurred around 300 years before the expected takeoff in AD 1900 and it was not a transition from stagnation to growth but from growth to growth. This feature, as well as the diversion to a slower trajectory at the time of the claimed takeoff around AD 1900, is ignored in the Unified Growth Theory.

**Summary and conclusions**

Unified Growth Theory (Galor, 2005a, 2011) is convincingly contradicted by the economic growth in Latin America. Here again, the theory tells one story while the data present a diametrically different account.

The theory claims that the epoch of stagnation continued until around AD 1900. The data show no convincing evidence of the existence of the epoch of stagnation. On the contrary they show that the economic growth was hyperbolic at least from AD 1600 and possibly even earlier. The theory claims a dramatic *takeoff* from stagnation to growth at the end of the alleged regime of stagnation. In contrast, the data show that at the time of the expected takeoff the economic growth started to be diverted to a *slower* trajectory. Galor's takeoff is replaced by a slower growth. Galor's "remarkable" and "stunning" escape from Malthusian trap (Galor, 2005a, pp. 177, 220) never happened because there was no trap.

Unified Growth Theory is repeatedly contradicted by data describing economic growth in Africa, Asia, former USSR, Western Europe and Eastern Europe (Nielsen, 2014, 2015b, 2015c, 2015d, 2015e), by the data describing the world economic growth (Nielsen, 2014, 2015a) and now also by the data describing economic growth in Latin America.

Unified Growth theory was developed over a long time of about 20 years (Baum, 2011) but it contains so many fundamental errors that it will take a long time and many more publications to correct them. In due time we shall demonstrate that this theory is contradicted not only by the GDP data but also by the GDP/cap data, as it has been already demonstrated by the discussion of the world economic growth (Nielsen, 2015a). We shall show that this theory is contradicted not only by the regional and global economic growth but also by national economic growth. In particular we shall show that it is contradicted by the economic growth in the UK, the very centre of the Industrial Revolution where the Unified Growth Theory should have the strongest support. We shall demonstrate that the postulate of the differential takeoffs and the postulate of the great divergence are also contradicted by data. We shall show that these two postulates are also based on the incorrect interpretation of the mathematical properties of hyperbolic distributions. Furthermore, we shall demonstrate that Galor's repeated interpretation of growth rates is incorrect.

Unified Growth Theory needs to be thoroughly revised but most likely it has to be replaced by a theory based on a scientific analysis of data because in its present form it contains far too much incorrect and misleading information. If economic growth research is to be treated as science, then this theory, which is full of unsupported stories, has to be either thoroughly and promptly revised or rejected and replaced by a new theory based on a scientific analysis of data.

In science, theories come and go and there is nothing unusual about it. This is the way science works and that is why science is a self-correcting discipline. Science has no room for dogmas, which have to be accepted by faith and emotionally defended. Incorrect interpretations have to be abandoned and replaced by interpretations supported by the rigorous analysis of data.

Pereira, E. (2011). Developing Countries Will Lead Global Growth in 2011, Says World Bank. http://www.forbes.com/sites/evapereira/2011/01/12/developing-countries-will-lead-global-growth-in-2011-says-world-bank/

Snowdon, B. & Galor, O. (2008). Towards a Unified Theory of Economic Growth. *World Economics*, *9*, 97-151.8